# A Novel Fast Response and Radiation-resistant Scintillator Detector for Beam Loss Monitor

**Y. Ji, Z. Tang[*], C. Li, X. Li and M. Shao**

  *Department of Modern Physics, University of Science and Technology of China (USTC).*
  *Hefei 230026, China*
  *State Key Laboratory of Particle Detection and Electronics,*
  *USTC-IHEP, China*
  *E-mail*: zbtang@ustc.edu.cn

ABSTRACT: At high luminosity area, beam loss monitor with fast response and high radiation resistance is crucial for accelerator operation. In this article, we report the design and test results of a fast response and radiation-resistant scintillator detector as the beam loss monitor for high luminosity collider, especially at low energy region such as RFQ. The detector is consisted of a 2 cm × 2 cm × 0.5 cm LYSO crystal readout by a 6 mm × 6 mm Silicon photomultiplier. Test results from various radioactive sources show that the detector has good sensitivity to photons from tens of keV to several MeV with good linearity and energy resolution (23% for 60 keV γ-ray). For field test, two such detectors are installed outside of the vacuum chamber shell of an 800 MeV electron storage ring. The details of the test and results are introduced.

---

[*] Corresponding author.

# Contents

## 1. Introduction

Beam loss has always been an essential problem to the protection of accelerator system, especially to high luminosity colliders. Due to the high intensity, even a tiny fraction loss of the beam may cause cooling issue and serious damage to accelerator components and radiation sensitive equipments. The primary goal of a beam loss monitor (BLM) system is to identify the loss level and, if possible, the spatial and time profile of the loss, to protect the equipments from uncontrolled loss (i.e., the uncontrolled beam loss as soon as possible and shuts down the accelerator in microsecond level). It also provides diagnostic information of the beam. For low energy (<5 MeV) but high intensity proton accelerator such as RFQ, BLM detector should be sensitive to photon, due to the relatively high yield of γ-ray and low yields of neutron and charged particle as seen from the simulation results shown in Fig. 1 [1].

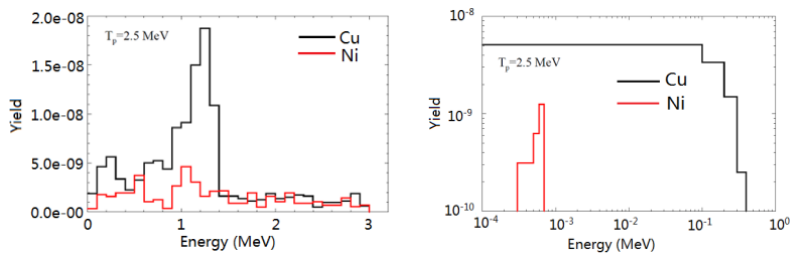

**Figure 1**: Secondary products by 2.5MeV proton beam hitting the Cu/Ni shell. The left one is γ-ray spectrum and the right one is neutron spectrum. The figures are taken from [1].

A good BLM should have the following charateristics: sensitivity to secondary particles ( here is photon), insensitivity to magnetic field, fast response, low cost, and high radiation-resistance. Ionization chamber and BF$_3$ counter tube are widely used as beam loss detectors, but they couldn't response in such a short time (μs). PIN diode is fast enough for beam loss monitor. However, it is usually not sensitive to γ-ray especially at the energy level of MeV. Plastic scintillator and liquid detector have relative good sensitivity to photon and fast response, but they are not resistant to the high intensity radiation. Usually, they use photomultiplier tube (PMT), the traditional photoelectric sensor, as read out. The PMTs are sensitive to the electromagnetic field, which limits the application of scintillator detector around electromagnetic coils. To satisfy all of the requirements mentioned above, a novel design of BLM detector is proposed, using LYSO crystal scintillator coupled with Silicon photomultiplier (SiPM). LYSO is a novel crystal scintillator. It features high light yield, short decay time, relatively large density and very high radiation resistance (100 Mrad for γ-ray [3]). SiPM is one of the most widely used photosensors. It has a relatively low working voltage (usually less than 100 V) with a gain at a level of $10^6$, which is comparable to PMT. It is compact, low cost, insensitive to magnetic field, and with a good resolution and a broad spectra response. The detector is designed for the RFQ section of the high intensity proton accelerator. Thanks to its good sensitivity and compact structure, it can also be applied to other kinds of low energy proton accelerator and general electron accelerators.

## 2. Detector design and properties

The detector is consisted of a 2 cm × 2 cm × 0.5 cm LYSO crystal as the scintillator and SiPM as readout, with high resistance magnetic field and a compact structure to fit in the limited volume in the collider. The surfaces of the crystal are wrapped by Tyvek, leaving a 6 mm × 6 mm window to couple with the SiPM.

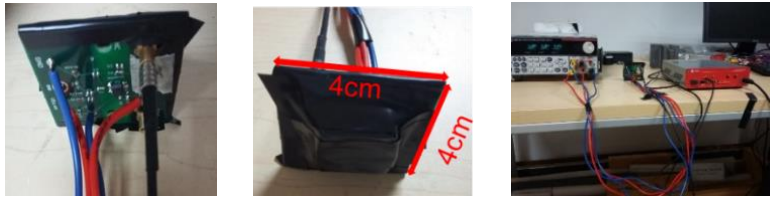

**Figure 2:** Overview of the detector

The SiPM used is SensL MicroFC 60035. Figure 3 shows the schematic of the read out circuit. The signal is collected by a capacitor and amplified by a factor of about 10 with high bandwidth. The waveforms of the detector output signals are recorded by a digitizer (CAEN DT5720). A typical waveform is shown in Fig. 4(a). The rise time of the detector signals has a mean of 47ns and variance (σ) of 3.6 ns. To obtain the gain of the SiPM, we analyze the dark-noise signals after the γ-ray signals as shown in Fig. 4(c). The charge distribution of the dark noises are shown in Fig. 5(a) and fit with a multi-Gaussian function. The mean values of the peaks as a function of number of photoelectrons are shown in Fig. 5(b). The gain is determined to be about $2.1 \times 10^6$ through the linear fit shown in Fig. 5(b).

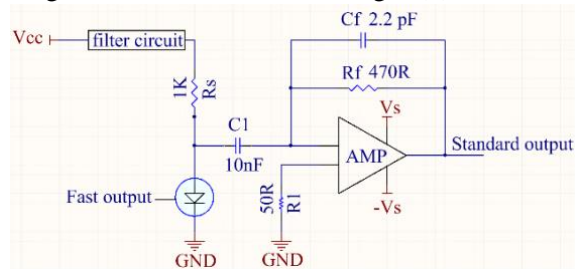

**Figure 3.** The schematic of the SiPM read out circuit

The detector is tested using various radioactive sources ($^{241}$Am, $^{22}$Na, $^{137}$Cs and $^{60}$Co). The energy spectra are shown in the Fig. 6. Fig. 6(a) shows that the detector could distinguish $^{241}$Am's 59.6 keV γ-ray from the pedestal clearly, which means the detector have high enough signal-to-noise ratio to detect low-energy photons. As the γ-ray energy goes higher, Compton Effect becomes significant. The spectra of $^{137}$Cs and $^{22}$Na are fitted with the combination of Fermi-Dirac function and Gaussian function. The Fermi-Dirac function is used to describe the Compton edge [3] and the Gaussian function is used to describe the full energy peak. The energy resolution of $^{137}$Cs's 0.662 MeV γ-ray is 13.5%. Here the energy resolution is defined as σ/$E$, where E is the energy of the full energy peak and σ is the standard deviation. In Fig. 6(c), the energy of $^{60}$Co's 1170 keV and 1330 keV γ-ray are too close to be distinguished. Since the two decay's branching ratios are almost the same, so two Gaussian function with the same amplitude and σ plus a Fermi-Dirac function are used to fit the spectrum. The energy calibration result is shown in Fig. 7. The detector achieves a good linearity at the energy level of MeV. Comparing with Fig. 5(b), the light yield of the detector is calculated to be 0.75 photoelectron per keV deposited energy by photons or electrons.

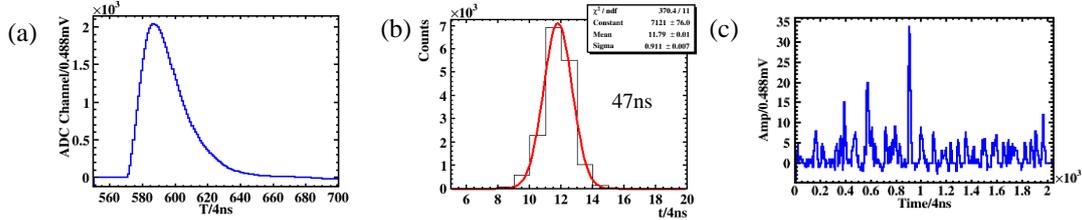

**Figure 4:** (a) A typical signal waveform of the γ-ray signal. (b) The distribution of the signal rise time. (c) Waveform of dark noises.

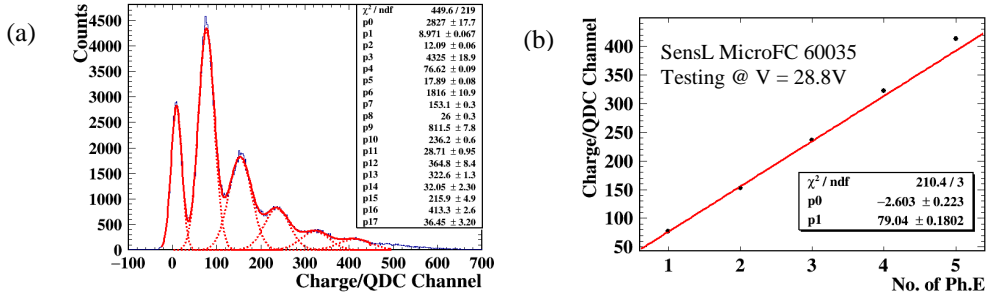

**Figure 5.** (a) Charge distribution of the dark noises, the solid line shows a fit to multi-Gaussian function. (b) Mean charge of the dark noises as a function of number-of-photoelectrons.

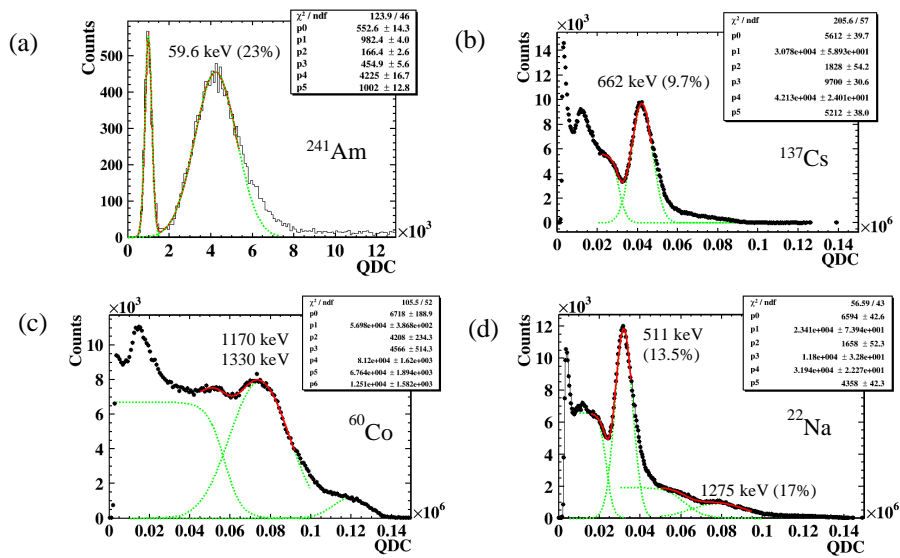

**Figure 6.** Energy spectra of different radioactive sources

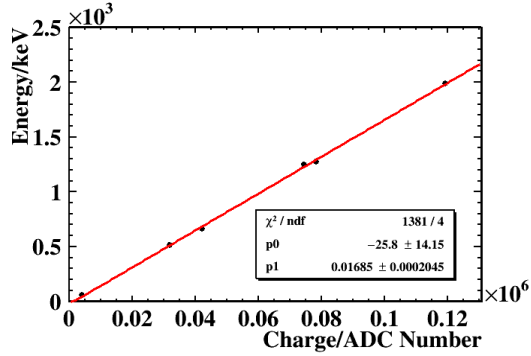

**Figure 7.** Energy calibration

## 3. Beam test at NSRL

### 3.1 Experiment setup

The detector is tested at the National Synchrotron Radiation Laboratory (NSRL), Hefei, China. Two detectors are installed outside the shell of vacuum chamber of an 800 MeV electron storage ring. One detector is parallel to the beam and the other is perpendicular. The position of the detectors along the accelerator tube are shown in Fig. 8. The escaping electrons hit the vacuum chamber shell at a very small angle (~2 degree), and generate electromagnetic showers. The detectors detect the secondary photons and electrons outside the vacuum chamber to monitor the beam loss level. The block diagram of the experiment is shown in Fig. 9.

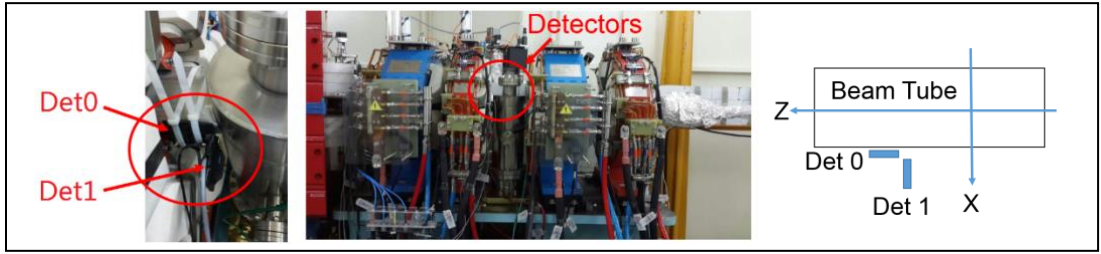

**Figure 8.** The installation position of the detectors. Rightmost plot is the verticle view of the beam tube, and the blue filled rectangles depict the two detectors installed.

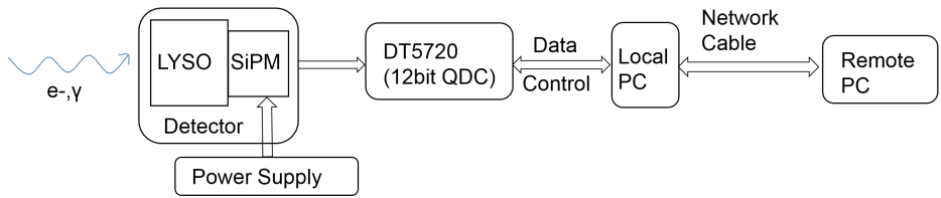

**Figure 9.** The block diagram of the experiment.

### 3.2 Beam test results

Fig. 10 shows the trigger rate of our two detectors as a function of time. The beam current are also shown for comparison. Beam loss rate has positive correlation with beam current. However, if there are only several bunches escaping from their determined tracks, the beam current won't drop apparently. The beam loss detector can provide us more detailed diagnostic information than the beam current. When unintended beam loss happens, a sharp peak can be observed from the time spectrum of the detector trigger frequency, shown in region A in Fig. 10,

– 4 –

and a zooming-in in Fig. 11(a). When the beam current drops to a certain threshold, the bunch of electrons will be injected with 1Hz frequency. A small bump can be observed at the time of injection (the region B in Fig. 10). Zooming in the region B, 1 Hz pulse-like beam loss during the injection can be clearly seen in Fig. 11(b). Fig. 11(c) is the further zooming-in of one of the pulses in Fig. 11(b). The time structure of the beam loss spectrum indicate that the detectors can response to the beam loss within 50µs, which is dominantly contributed by the rise time of the injection and/or beam loss instead of the detector response time. The detected energy spectrum of the beam loss is drawn in Fig. 13. It is discussed in the following section and compared with the simulation results.

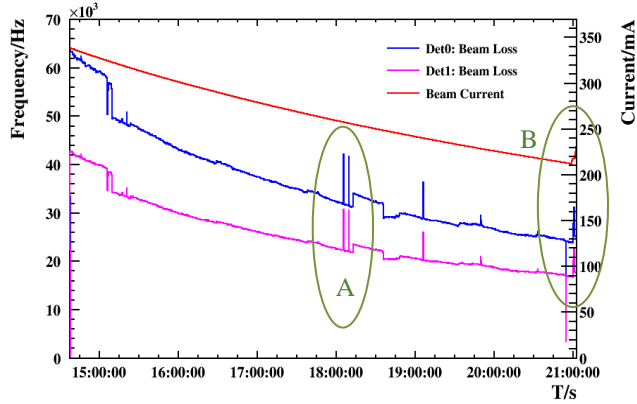

**Figure 10.** The detectors trigger frequency and the beam current versus time.

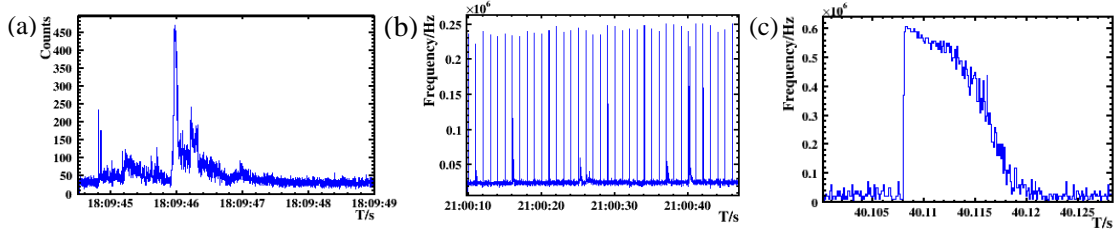

**Figure 11.** (a) The zooming-in of the region A in Fig. 10. (b) The zooming-in of the region B in Fig. 10. (c) The zooming-in of one of the pulses in Pic. (b).

### 3.3 Geant4 simulation results

Monte-Carlo simulation is carrying out by using Geant4 toolkit. The beam tube is constructed by Geant4 according to the real structure (Fig. 12(a)). The detectors are placed at the same position as in Fig. 8. In the simulation, 800 MeV electrons hit the shell of the beam tube at an angle of 2 degree with respected to the the central axis of the beam tube, as illustrated in Fig. 12(b). The spectra of energy deposition from different particles are shown in Fig. 13(a). When the deposited energy is below 10MeV, the simulation results can describe the experimental results. When the deposited energy goes above 10MeV, the digitizer and the SiPM are gradually saturated. Fig. 13(b) shows the different particles' contributions to the energy spectrum. It can be inferred that there are more than one secondary particles hitting the detector in several events. The energy of secondary photon deposited in the detectors is mainly below 2 MeV. For energy above 2 MeV, the spectrum is dominated by electrons and positrons.

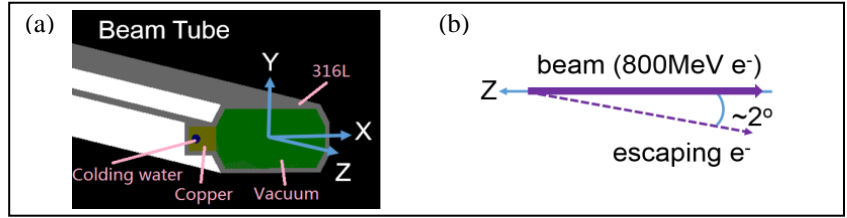

**Figure 12.** The left one shows the structure and the material of the beam tube. The right one shows the direction and energy of the incident electrons.

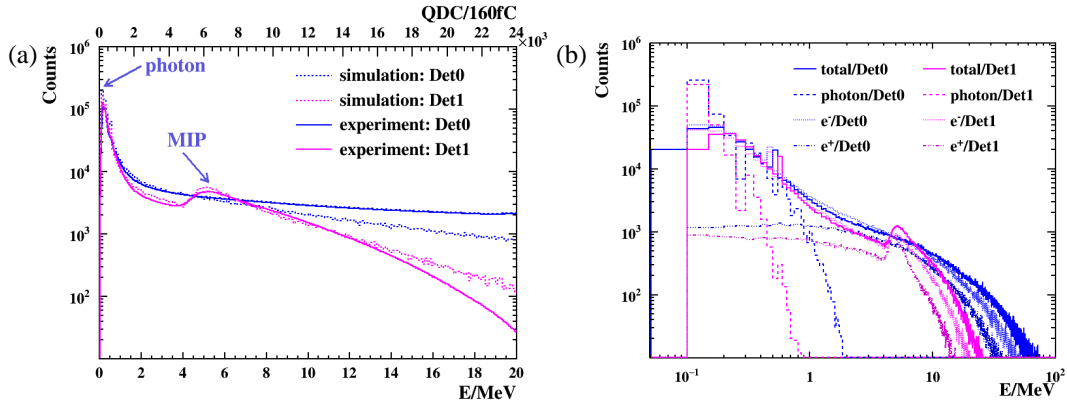

**Figure 13.** (a) The energy spectrum of the beam loss. (b) The simulated spectra of energy deposited in the detectors contributed by different kinds of secondary particles.

## 4. Summary

The novel BLM detector consisting of LYSO crystal coupled with SiPM features fast response, high radiation resistance, compact volume, as well as insensitivity to magnetic field. It also has a large dynamic energy range, keeping good sensitivity to photon and charged particles from tens of keV to several MeV with good linearity. Being sensitive to the beam loss, it could provide more detailed information than beam current for furthermore diagnosis of the beam. In the field test, the detector keeps a good performance when exposed to high background of several MeV to hundreds of MeV electrons for two months. We are planning to test it at the high intensity proton accelerator of the Accelerator Driven Sub-critical System of China (C-ADS) and China Spallation Neutron Source (CNSR) in the future.

## References

[1] N. Mokhov, I. Rakhno. *Neutron and Photon Production by Low Energy Protons*. Fermilab-FN-0909-APC, December 2010.

[2] Yang, Fan, Liyuan Zhang, and Ren-Yuan Zhu. *Gamma-Ray Induced Radiation Damage Up to 340 Mrad in Various Scintillation Crystals*. IEEE Transactions on Nuclear Science 63.2 (2016) 612-619.

[3] Baccaro S et al, *Precise determination of the light yield of scintillating crystals*, Nucl. Instrum. Methods A 385.1 (1997) 69.